# Tunable ferromagnetism at non-integer filling of a moiré superlattice

Guorui Chen[1+], Aaron L. Sharpe[2,3+], Eli J. Fox[3,4+], Shaoxin Wang[1], Bosai Lyu[5,6], Lili Jiang[1], Hongyuan Li[1,7], Kenji Watanabe[8], Takashi Taniguchi[9], Michael F. Crommie[1,7,10], M. A. Kastner[3,4,11], Zhiwen Shi[5,6], David Goldhaber-Gordon[3,4*], Yuanbo Zhang[6,12,13*], Feng Wang[1,9,14*]

[1]Department of Physics, University of California at Berkeley, Berkeley, CA 94720, USA.

[2]Department of Applied Physics, Stanford University, Stanford, CA 94305, USA.

[3]Stanford Institute for Materials and Energy Sciences, SLAC National Accelerator Laboratory, Menlo Park, CA 94025, USA.

[4]Department of Physics, Stanford University, Stanford, CA 94305, USA.

[5]Key Laboratory of Artificial Structures and Quantum Control (Ministry of Education), Shenyang National Laboratory for Materials Science, School of Physics and Astronomy, Shanghai Jiao Tong University, Shanghai 200240, China.

[6]Collaborative Innovation Center of Advanced Microstructures, Nanjing, Jiangsu 210093, China.

[7]Materials Science Division, Lawrence Berkeley National Laboratory, Berkeley, CA 94720, USA.

[8]Research Center for Functional Materials, National Institute for Materials Science, 1-1 Namiki, Tsukuba 305-0044, Japan.

[9]International Center for Materials Nanoarchitectonics, National Institute for Materials Science, 1-1 Namiki, Tsukuba 305-0044, Japan.

[10]Kavli Energy NanoSciences Institute at the University of California, Berkeley and the Lawrence Berkeley National Laboratory, Berkeley, CA 94720, USA.

[11]Department of Physics, Massachusetts Institute of Technology, Cambridge, MA 02139, USA.

[12]State Key Laboratory of Surface Physics and Department of Physics, Fudan University, Shanghai 200433, China.

[13]Institute for Nanoelectronic Devices and Quantum Computing, Fudan University, Shanghai 200433, China.

[14]Kavli Energy NanoSciences Institute at the University of California, Berkeley and the Lawrence Berkeley National Laboratory, Berkeley, CA 94720, USA.

[+]These authors contributed equally to this work.
*Correspondence to: fengwang76@berkeley.edu, zhyb@fudan.edu.cn, goldhaber-gordon@stanford.edu

The flat bands resulting from moiré superlattices in magic-angle twisted bilayer graphene (MATBG) and ABC-trilayer graphene aligned with hexagonal boron nitride (ABC-TLG/hBN) have been shown to give rise to fascinating correlated electron phenomena such as correlated insulators(*1*, *2*) and superconductivity(*3*, *4*). More recently, orbital magnetism associated with correlated Chern insulators was found in this class of layered structures centered at integer multiples of $n_0$, the density corresponding to one electron per moiré superlattice unit cell(*5–7*). Here we report the experimental observation of ferromagnetism at fractional filling of a flat Chern band in an ABC-TLG/hBN moiré superlattice. The ferromagnetic state exhibits prominent ferromagnetic hysteresis behavior with large anomalous Hall resistivity in a broad region of densities, centered in the valence miniband at $n = -2.3\ n_0$. This ferromagnetism depends very sensitively on the control parameters in the moiré system: not only the magnitude of the anomalous Hall signal, but also the sign of the hysteretic ferromagnetic response can be modulated by tuning the carrier density and displacement field. Our discovery of electrically tunable ferromagnetism in a moiré Chern band at non-integer filling highlights the opportunities for exploring new correlated ferromagnetic states in moiré heterostructures.

Moiré superlattices in van der Waals heterostructures have emerged as a powerful platform for exploring novel correlated electron physics. The electronic structure of the moiré miniband can be conveniently controlled by the stacking sequence of layers and by electric fields applied normal to the layers(*8*), and the moiré bandwidth can be made small enough that electron correlations predominate(*9, 10*). In addition, the electron density can be tuned across an entire band using gates, something that is much more difficult in more conventional correlated electron materials. In graphene-based moiré superlattices, such control has enabled discovery of correlated insulators, superconductivity, and Chern insulators with orbital ferromagnetism(*1–7, 11–15*); Mott insulators and generalized Wigner crystals have been reported in transition metal dichalcogenide moiré superlattices(*16–18*).

Here we report the observation of ferromagnetism at non-integer filling of a moiré superlattice band of ABC-TLG/hBN. We perform magneto-transport measurements to map out the phase diagram of an ABC-TLG/hBN heterostructure as a function of the carrier density $n$, the vertical displacement field $D$, and the vertical magnetic field $B$. In addition to the orbital ferromagnetism and a Chern insulator state, we previously identified(*6*), at $n = -n_0$ (one hole per moiré unit cell), we discover another region of ferromagnetism when the topological Chern band is partially filled between $n = -2.1\ n_0$ and $-2.7\ n_0$ and $D$ ranges from -0.45 V/nm to -0.55 V/nm. The ferromagnetic state exhibits hysteresis and shows a zero-field anomalous Hall (AH) signal as large as -4.8 k$\Omega$ at $n = -2.3\ n_0$ and $D = -0.48$ V/nm. The large AH signal is accompanied by a peak in the longitudinal resistivity as a function of $n$. In addition, the longitudinal resistivity of the magnetic states increase slightly when reducing temperature, suggesting that the states showing ferromagnetism are insulating. We further find that the ferromagnetism depends sensitively on the carrier density and vertical displacement field: not only the magnitude of the AH signal, but also the sign of the ferromagnetic hysteresis can be controlled by the carrier density and displacement field.

The ABC-TLG/hBN moiré heterostructures are fabricated following the method described in (*6*). We first identify the ABC-TLG domain using scanning near-field infrared nanoscopy and then isolate the ABC domain from adjacent ABA regions using atomic force microscope cutting(*19, 20*). The isolated ABC-TLG is encapsulated in exfoliated hBN layers, where one hBN layer is aligned with the ABC-TLG to form the moiré superlattice(*21*). For magneto-transport measurements, the ABC-TLG/hBN heterostructures are fabricated into a Hall bar geometry with a metal top gate and a doped silicon bottom gate. During fabrication, the ABC-TLG often relaxes to the lower energy ABA state, but after fabrication we find that the ABC stacking is stable. An optical image of the Hall bar device and the Hall measurement configuration are displayed in Fig. 1A. Figure 1B shows a schematic side-view of the dual-gated device. The top and bottom gate voltages ($V_t$ and $V_b$) allow us to independently control the carrier density and the miniband bandwidth and topology of the ABC-TLG/hBN heterostructure: the doping relative to the charge neutrality point is set by $n = (D_b - D_t)/e$, and the miniband bandwidth and topology are controlled by the applied vertical displacement field $D = (D_b + D_t)/2$. Here $D_b = +\varepsilon_b(V_b - V_b^0)/d_b$ and $D_t = -\varepsilon_t(V_t - V_t^0)/d_t$ are the vertical displacement field below and above the ABC-TLG/hBN moiré superlattice, respectively, $\varepsilon_{b(t)}$ and $d_{b(t)}$ are the dielectric constant and thickness of the bottom (top) dielectric layers, and $V_{b(t)}^0$ is the effective offset in the bottom (top) gate voltage caused by environment-induced carrier doping. The longitudinal resistance is $R_{xx} = \frac{V_{xx}}{I}$ and Hall resistance is $R_{yx} = \frac{V_{yx}}{I}$. All measurements are performed at 0.07 K and with an AC excitation current $I$ = 0.5 nA rms unless otherwise noted.

The bandwidth and topology of the moiré minibands in ABC-TLG/hBN can be controlled by the vertical displacement field *D*. Figures 1C, D and E display the calculated single-particle bandstructure in ABC-TLG/hBN for the K valley at *D* = 0.5 V/nm, 0 V/nm, -0.5 V/nm, respectively(*6*). At *D* = 0.5 V/nm a finite bandgap separates the conduction and valence miniband. The highest-energy valence miniband is flat with a bandwidth as narrow as 20 meV, and it is topologically trivial with *C* = 0. At *D* = 0 V/nm, the conduction and valence moiré minibands become broader and overlapping so that the band gap closes. At *D* = -0.5 V/nm, a bandgap reopens, with band topology changed relative to positive *D* field. Specifically, the highest-energy valence miniband becomes topologically nontrivial with a finite Chern number and a narrow bandwidth of 20 meV. The electronic states in the K' valley are time reversed pairs of those in the K valley, and the topological minibands have opposite Chern numbers. The valence miniband, because it has a smaller bandwidth, should have stronger e-e interactions compared to the conduction miniband(*22–26*). We are therefore not surprised that the interesting phenomena reported here occur with hole, rather than electron, doping.

This electronic bandstructure of the moiré minibands can be conveniently tuned by the $D$ field in ABC-TLG/hBN heterostructure. Figure 1F displays the longitudinal resistance $R_{xx}$ as a function of band filling $n / n_0$ and displacement field $D$ in the device(*27*) ($n$ and $D$ are linearly transformed from the gate voltages, as shown in Fig. S1). In addition to the band insulating states (characterized by resistance peaks) at the charge neutral point ($n = 0$) and the fully filled point ($n = -4n_0$), correlated insulator states also emerge at 1/4 filling ($n = -n_0$) and 1/2 filling ($n = -2n_0$) of the hole minibands (i.e. one and two holes per moiré unit cell, respectively).(*2*) These correlated states appear at finite displacement field $|D|$, presumably because they require narrowing of the moiré minibands.

To explore ferromagnetism and evidence for band topology, we turn to magneto-transport measurements. Figure 2A and 2B show the longitudinal resistance at $B = 0$ T as a function of the band filling of topologically trivial and non-trivial valence minibands at $D = 0.4$V/nm and -0.47 V/nm, respectively. In the trivial flat band (at $D = 0.4$ V/nm), strong Mott insulator states are present at $n = -n_0$ and $n = -2 n_0$. In the topological flat band (at $D = -0.47$ V/nm), weaker resistance peaks are observed at $n = - n_0$, $n = -2 n_0$, and $n = -2.3 n_0$. Figure 2C and D display the corresponding two-dimensional plots of the Hall resistance $R_{yx}$ as a function of the hole doping and the perpendicular magnetic field $B$ (normal to ABC-TLG/hBN layers). In the trivial flat band (at $D = 0.4$ V/nm), the Hall resistance signal tends to be rather small for all hole doping (Fig. 2C) and no ferromagnetic signature is observed. (The large signals at $n = - n_0$ and $n = -2 n_0$ fillings are artifacts coming from crosstalk from the large $R_{xx}$ of the correlated insulator states; they do not change sign when the magnetic field is reversed.) With displacement field reversed $D = -0.47$ V/nm, the behavior is strikingly different: at carrier densities around $n = -n_0$ and $n = -2.3n_0$, the Hall resistance $R_{yx}$ is very large at weak magnetic fields and persists down to zero magnetic field. $R_{yx}$ switches sign abruptly at a magnetic field close to, but away from, zero (Fig. 2D; field is swept from positive to negative). Such behavior is characteristic of a ferromagnetic state. Indeed, the state at $n = -n_0$ has been previously(*6*) shown to be a $C = 2$ Chern insulator state.(*28*) The Chern insulator at $n = - n_0$ can be understood as arising from a single fully-filled valley-polarized Chern band(*26*). The state around $n = -2.3 n_0$, in contrast, is a completely new magnetic state that has not been predicted and emerges, surprisingly, at non-integer filling of the moiré miniband.

To establish the region of parameter space where the new magnetic state exists, we applied $B = 0.2$ T to favor polarization, and track $R_{yx}$ as a function of $n$ and $D$ (Fig. 3A). In an oval region of parameter space centered around $n = -2.3\ n_0$ and $D = -0.48$ V/nm (outlined by the dashed line in Fig. 3A, which is discussed in more detail below), $R_{yx}$ is large and negative. Figure. 3B shows the corresponding two-dimensional plots of $R_{xx}$ as a function of $n / n_0$ and $D$ at a weak magnetic field $B = 0.2$ T. This oval region also shows enhanced $R_{xx}$ compared with the surrounding region, as shown in Fig. 3B. This suggests that the ferromagnetic phase in this range of parameters is always insulating.

Next we examine in detail the $B$ field and temperature dependence of the new magnetic state at which the AH has the largest magnitude ($n = -2.3n_0$ and $D = -0.48$ V/nm). We measure the anti-symmetrized Hall resistivity $\rho_{yx}(B, \text{up}) = \frac{R_{yx}(B,\text{up}) - R_{yx}(-B,\text{down})}{2}$, $\rho_{yx}(B, \text{down}) = \frac{R_{yx}(B,\text{down}) - R_{yx}(-B,\text{up})}{2}$ and symmetrized longitudinal resistivity $\rho_{xx}(B, \text{up}) = \frac{R_{xx}(B,\text{up}) + R_{xx}(-B,\text{down})}{2} \cdot \frac{W}{L}$, $\rho_{xx}(B, \text{down}) = \frac{R_{xx}(B,\text{down}) + R_{xx}(-B,\text{up})}{2} \cdot \frac{W}{L}$, where up/down indicates the direction in which the magnetic field was being swept, the channel width $W = 1$ μm, and the channel length $L = 4$ μm. (Throughout this paper, we use $\rho_{yx}$ when presenting AH signals as functions of the magnetic field to eliminate longitudinal resistance contributions, but we use the directly measured $R_{yx}$ when presenting color maps of the dependence of Hall resistance on $n$ and $D$.) When $B$ is swept over the narrow range between -0.1 to 0.1 Tesla, the Hall resistivity $\rho_{yx}$ shows a clear AH signal with strong ferromagnetic hysteresis (Fig. 3C). The AH signal reaches $\rho_{yx}^{AH} = -4.8$ kΩ with a coercive field of $B_c = 0.04$ T at the base temperature of $T = 0.07$ K. $\rho_{yx}^{AH}$ decreases monotonically with increasing temperature, reaching almost zero at $T = 1.6$ K (Fig. 3E). We note that $\rho_{yx}^{AH}$ at $n = -2.3n_0$ exhibits the opposite sign compared to that of the Chern insulator at $n = -n_0$ (*27*)(Fig. S2 shows a direct comparison between these two magnetic states).

Figure 3D shows that the overall longitudinal resistivity of the magnetic state increases as temperature is lowered. Figure 3E and F display the AH signal $\rho_{yx}^{AH}$ and longitudinal conductivity $\sigma_{xx}$ (derived from the measured resistivities) as functions of the temperature at $B = 0$. As the temperature is lowered, the ferromagnetic state strengthens, as evinced by increasing $\rho_{yx}^{AH}$ and the conductivity $\sigma_{xx}$ decreases. The temperature dependences of $\rho_{xx}$ and $\sigma_{xx}$ suggest that the ferromagnetic state is insulating. However, we do not observe well-defined Arrhenius behavior (see plot of log $\sigma_{xx}$ versus $1/T$ in Fig. S5C).

Next we examine the evolution of the AH signal as a function of $D$ and $n$. We sweep the field up and down over a small range of magnetic fields and measure $\rho_{yx}$. Figure 4A shows the $D$-dependence at fixed doping of $n = -2.5\ n_0$. Beginning at $D = -0.49$ V/nm, where there is no nonlinearity, the hallmark of AH, the AH grows with less negative $D$ as one approaches the center of the oval. Upon tuning away from the center of the oval, the AH signal decreases quickly and becomes zero at $D = -0.38$ V/nm. Surprisingly, the AH reappears at $D = -0.37$ V/nm, accompanied by a switch in sign of the hysteresis loop. It then decreases and becomes very small again for $D \gtrsim -0.35$ V/nm.

The magnitude and sign of the ferromagnetic hysteresis also depends sensitively on the carrier density $n$ for a fixed $D = -0.39$ V/nm (Fig. 4B). At this displacement field, the AH signal reaches a first maximum at $n = -2.48\ n_0$. With less negative $n$, the AH signal decreases quickly. Then the AH signal changes sign, and although the hysteresis disappears near $n = -2.31\ n_0$, the non-linear component, which is the best measure of the AH signal, continues to grow as n is made less negative. Additional data of this type are shown in the supplemental material, showing that the AH signal vanishes before $n$ reaches $-2n_0$.

To visualize the dramatic change of the ferromagnetic state with $n$ and $D$, we indicate the parameter space with negative and positive $\rho_{yx}^{AH}$ in Fig. 4C with blue and red squares, respectively, where the intensity of the color indicates the magnitude of the anomalous Hall signal. Values of ($n$, $D$) where AH is zero are indicated by circles. A variety of additional magnetic field sweeps for various values of $n$ and $D$ are shown in Fig. S4 for the same sample studied in Fig. 4 and in Fig. S7 for a second sample(*27*). We note that the hysteresis loops of the magnetic state can be rather complicated close to the oval boundaries, at which the AH vanishes. Some hysteresis loops show multiple intermediate jumps in $\rho_{yx}$ ($R_{yx}$), and some hysteresis loops appear at non-zero magnetic fields (Fig. 4 and Fig. S4).

We have shown that ferromagnetism can emerge not only at integer filling of a miniband in an ABC-TLG/hBN moiré superlattice(*6*) but also in a region centered about non-integer filling, behavior which is unique among the reported magnetic states in moiré systems. As with AH in other moiré systems, the system appears to be insulating at zero field and the magnitude of the AH is large, indicating that it is unlikely that the AH arises from some scattering mechanism as in ferromagnetic metals. We therefore assume that, while the ferromagnetic state at non-integer filling in ABC-TLG/hBN as yet exhibits significant longitudinal resistance and no quantization of AH, the large AH arises, as it does in the other systems, from a combination of band topology and e-e interactions. Interactions break valley/spin degeneracy, and give a gap between the bands with opposite Chern numbers. The observed magnetic state appears only when the valence miniband has been tuned by the applied displacement field to be topologically non-trivial, suggesting the importance of band topology for realizing the magnetic state.

The ferromagnetic hysteresis in Hall resistance at non-integer filling can change sign with field effect control of $D$ and $n$, perhaps indicating that different ferromagnetic states can be stabilized in different regions of parameter space(*29*). Similar behavior has been observed in a second device(*27*) (See Fig. S6-8). This reversal of the sign of AH as a function of carrier density is similar to that recently reported for ferromagnetism centered at $n = 3n_0$ in twisted monolayer-bilayer graphene(*30*). These authors report that the Hall resistance saturates near $h/2e^2$ on both sides of the sign reversal, where $h$ is Planck's constant and $e$ is the charge of an electron, while the longitudinal resistance $R_{xx} \leq 1$ k$\Omega$, suggesting that on the two sides of the sign reversal there are Chern insulators with Chern numbers $C = \pm 2$. The sign reversal of the AH with carrier density reported here can then be explained as follows: with an applied external magnetic field, we control the preferred magnetization direction, as the magnetic moment wants to align with the field. But Chern number is not simply related to magnetization: the magnetization of an orbital Chern insulator reverses as the chemical potential passes through the gap(*29*). Therefore the magnetization of each valley, and hence the magnetization for a Chern insulator of each Chern number, reverses as a function of carrier density, yielding the measured reversal of AH resistance. The sign of the AH is not reported to be sensitive to $D$ in twisted monolayer-bilayer graphene(*30*) as it is for the non-integer filling ferromagnetic state in the present work. However, it is predicted that the sign of the magnetization of an orbital Chern insulator of a given Chern number also depends on gap size, the bandwidth of the bands, and the valley splitting(*29*). Tuning $D$ strongly affects these parameters in ABC-TLG/hBN, therefore it is perhaps not surprising that, for fixed chemical potential, we see a sign flip in the measured AH resistance as a function of *D,* as well. Twisted monolayer-bilayer graphene has also been found to be magnetic in a region centered near $n = n_0$ but this state does not exhibit a sign change of AH as a function of carrier density(*30, 31*).

Thus both the large AH and its sign reversal likely arise from a correlation-induced gap between bands with opposite Chern numbers. Whereas it is straightforward to understand such a gap in twisted monolayer-bilayer graphene at integer filling as caused by electron-electron interactions simply by symmetry breaking of the spin/valley degrees of freedom(*29–31*), a gap at non-integer filling as in the ABC-trilayer graphene/hBN moiré system suggests a more complicated ground state.

**ACKNOWLEDGMENTS**

We acknowledge the help in the bandstructure calculations and helpful discussions with Ya-Hui Zhang and T. Senthil. **Funding:** G.C., H. L, M.C., and F.W. were supported as part of the Center for Novel Pathways to Quantum Coherence in Materials, an Energy Frontier Research Center funded by the U.S. Department of Energy, Office of Science, Basic Energy Sciences. A.S. acknowledges support from an ARCS Foundation Fellowship, a Ford Foundation Predoctoral Fellowship, and a National Science Foundation Graduate Research Fellowship. E.F., and D.G.-G.'s work on this project were supported by the U.S. Department of Energy, Office of Science, Basic Energy Sciences, Materials Sciences and Engineering Division, under Contract No. DE-AC02-76SF00515. Dilution fridge support: Low-temperature infrastructure and cryostat support were funded in part by the Gordon and Betty Moore Foundation through Grant No. GBMF3429. Part of the sample fabrication was conducted at Nano-fabrication Laboratory at Fudan University. Y.Z. acknowledges financial support from National Key Research Program of China (grant nos. 2016YFA0300703, 2018YFA0305600), NSF of China (grant nos. U1732274, 11527805 and 11421404), Shanghai Municipal Science and Technology Commission (grant nos. 18JC1410300 and 2019SHZDZX01) and Strategic Priority Research Program of Chinese Academy of Sciences (grant no. XDB30000000). K.W. and T.T. acknowledge support from the Elemental Strategy Initiative conducted by the MEXT, Japan, Grant Number JPMXP0112101001, JSPS KAKENHI Grant Number JP20H00354 and the CREST(JPMJCR15F3), JST. **Author contributions:** F.W., Y.Z. and D.G.-G. supervised the project. G.C. fabricated samples. G.C., A.S. and E.F. performed transport measurements. G.C., S.W., B.L., L.J., H.L. and Z.S. prepared trilayer graphene and performed near-field infrared and AFM measurements. K.W. and T.T. grew hBN single crystals. G.C., A.S., E.F., D.G.-G., Y.Z. and F.W. analyzed the data. G.C., A.S., M.A.K., D.G.-G and F.W. wrote the paper, with input from all authors. **Competing interests:** M.A.K. is chair of the Basic Energy Sciences Advisory Committee. Basic Energy Sciences provided funding for this work. **Data and materials availability:** The data that support the findings of this study are available from the corresponding authors upon reasonable request.

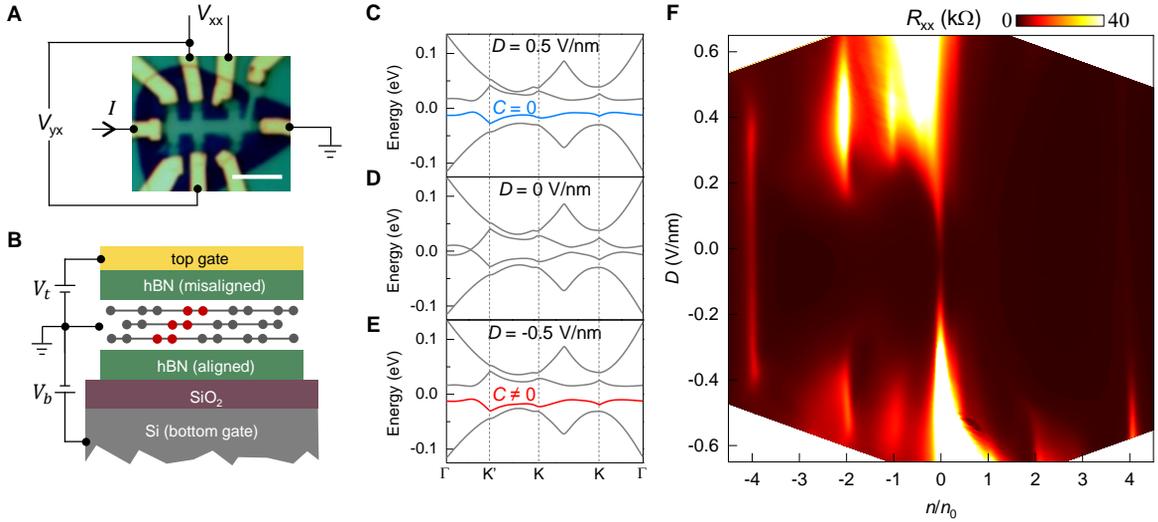

**Fig. 1. Device structure, band structure and longitudinal resistance of ABC-TLG/hBN.**
(**A**) Optical image of ABC-TLG/hBN Hall bar device and measurement configuration. White scale bar: 5 μm. (**B**) Schematic cross-sectional view of the dual-gated device shown in (A). The red atoms represent one unit cell of ABC-TLG. (**C**, **D** and **E**) Calculated single-particle band structure of ABC-TLG/hBN in the K valley for displacement fields $D = 0.5$, 0 and -0.5 V/nm. At $D = 0.5$ V/nm a finite bandgap separates the conduction and valence bands. The highest energy valence miniband is topologically trivial with $C = 0$. The bandgap closes at $D = 0$ V/nm. At $D = -0.5$ V/nm the band inverts to generate a new bandgap. The highest energy valence miniband becomes topologically nontrivial with a finite Chern number. The minibands of the K' valley are time-reversed pairs of the K valley minibands and have opposite Chern numbers. (**F**) Color plot of the measured longitudinal resistance $R_{xx} = V_{xx} / I$ as a function of band filling factor $n / n_0$ and vertical displacement field $D$ at $T = 1.5$ K. In addition to the band insulating states at the charge neutrality point ($n = 0$) and the fully filled point ($n = -4\, n_0$), correlated insulator states emerge at 1/4 filling ($n = -\, n_0$) and 1/2 filling ($n = -2\, n_0$) at finite $|D|$.

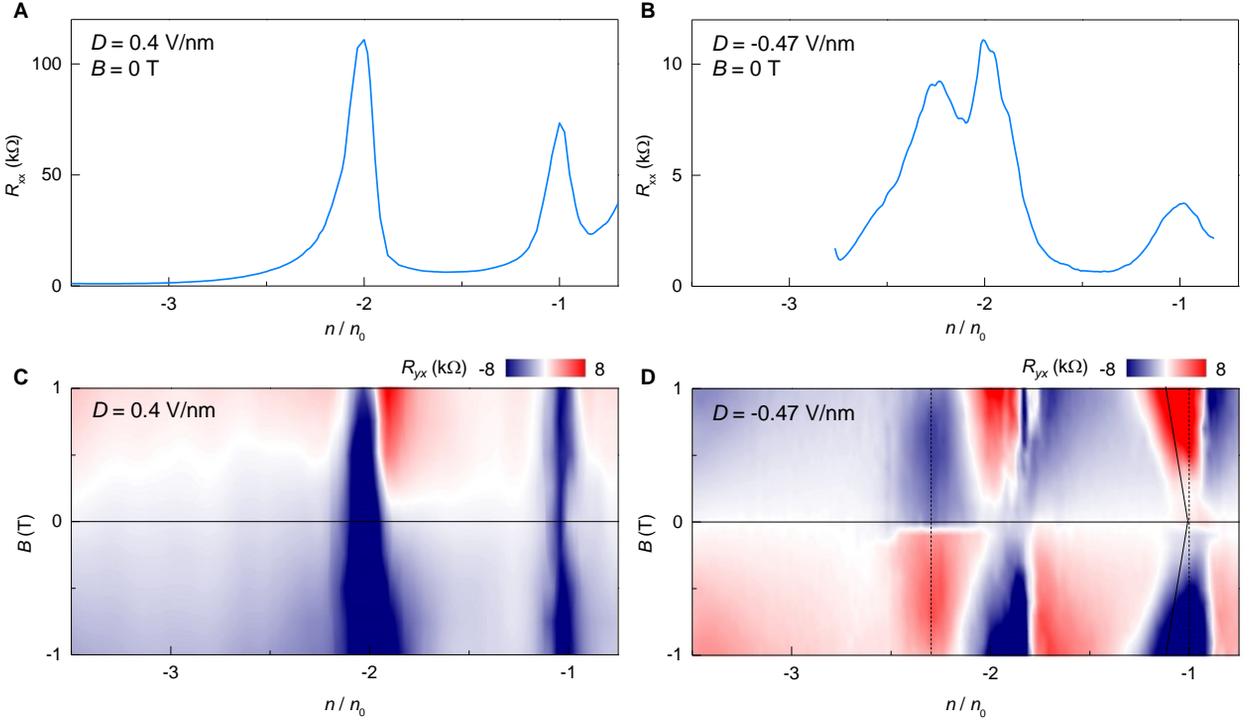

**Fig. 2. Longitudinal and Hall resistances at selected values of *D*.** (**A** and **B**) Longitudinal resistance as a function of band filling at $D = 0.4$ V/nm and $-0.47$ V/nm, respectively. (**C** and **D**) Corresponding color plots of the Hall resistance as a function of band filling and perpendicular magnetic field. For $D = 0.4$ V/nm in (C), the Hall resistance tends to be small for all doping and no ferromagnetic signature is observed. (The relatively large signals at $n = -n_0$ and $n = -2 n_0$ in (C) are artifacts coming from crosstalk from the large $R_{xx}$, evinced by the observation that they do not change sign when the magnetic field is reversed.) For $D = -0.47$ V/nm in (D), the Hall resistance $R_{yx}$ is very large at weak magnetic fields and can persist down to zero magnetic field at $n = -n_0$ and $n = -2.3 n_0$ (denoted by the two dashed vertical lines). The non-zero values of $R_{yx}$ at $B = 0$ represent AH signals from ferromagnetic states at $n = -n_0$ and $n = -2.3 n_0$. The state at $n = -n_0$ has been identified as a $C = 2$ Chern insulator state (two oblique solid lines) with orbital ferromagnetism (*6*). The AH at $n = -2.3 n_0$ results from a new magnetic state that emerges at non-integer filling of the topological hole miniband.

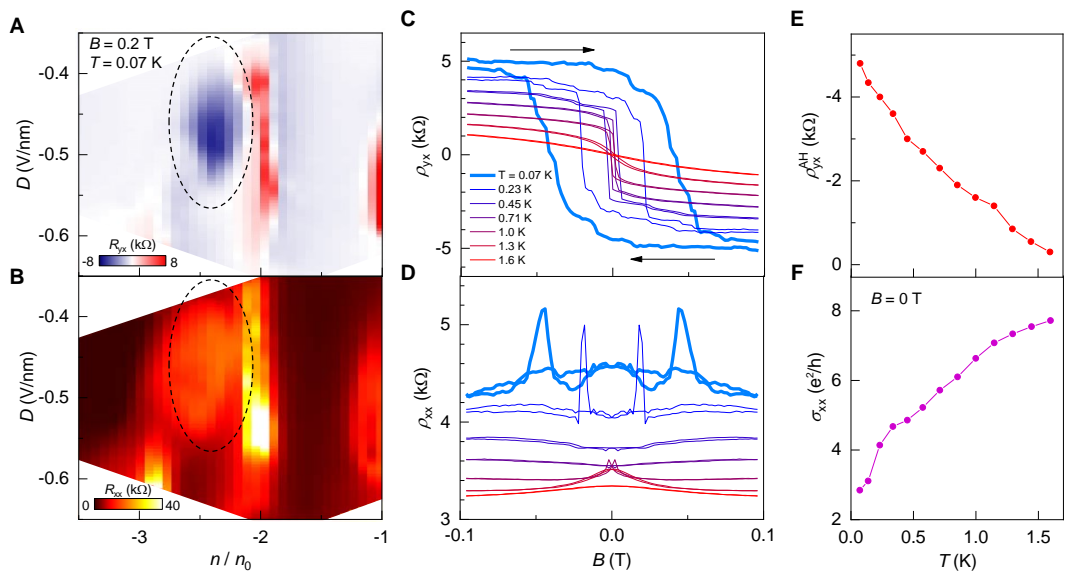

**Fig. 3. Anomalous Hall effect at non-integer filling.** (**A** and **B**) Color plots of $R_{yx}$ and $R_{xx}$ as functions of band filling and displacement field, respectively, for the valence miniband at $B = 0.2$ T and $T = 0.07$ K. The dashed oval outlines a region with large $R_{yx}$ signals at very small field between $n = -2.1\ n_0$ and $-2.6\ n_0$ and between $D = -0.4$ and $-0.53$ V/nm (more detail is given in Fig. 4). The oval region also shows enhanced $R_{xx}$ in (B), suggesting that the ferromagnetic phase is insulating. (**C** and **D**) Magnetic-field dependent antisymmetrized $\rho_{yx}$ and symmetrized $\rho_{xx}$ at different temperatures for the ferromagnetic state with largest AH at the center of the oval region ($n = -2.3\ n_0$ and $D = -0.48$ V/nm). The AH signal shows clear ferromagnetic hysteresis in (C), reaching $\rho_{yx}^{AH} = -4.8$ kΩ and a coercive field of $B_c = 0.04$ T at $T = 0.07$ K. (**E** and **F**) The evolution of $\rho_{yx}^{AH}$ and $\sigma_{xx}$ as a function of temperature at $n = -2.3\ n_0$ and $D = -0.48$ V/nm. The amplitude of $\rho_{yx}^{AH}$ decreases with increasing $T$, vanishing around 1.6 Kelvin, while $\sigma_{xx}$ increases with increasing $T$.

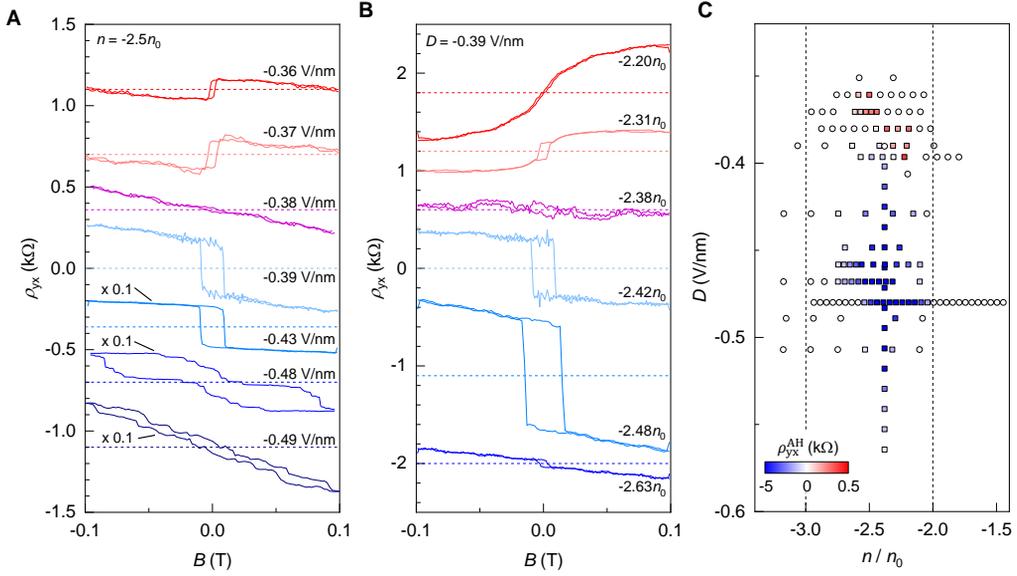

**Fig. 4. Tunable anomalous Hall effect at non-integer filling.** (**A**) The $D$-dependent ferromagnetism at $n = -2.5\, n_0$. The AH resistance is maximal at $D = -0.48$ V/nm, becomes zero at $D = -0.38$ V/nm, and then reappears with a sign change at $D = -0.37$ V/nm. The $\rho_{yx}$ signals at $D = -0.49$, $-0.48$ and $-0.43$ V/nm are manually multiplied by a factor of 0.1 for clarity. (**B**) The $n$-dependent ferromagnetism at a fixed $D = -0.39$ V/nm. The AH signal reaches a maximum at $n = -2.48\, n_0$, becomes nearly zero at $n = -2.38\, n_0$, and reappears with a sign change at $n = -2.31\, n_0$. Although the hysteresis is gone at $n = -2.2\, n_0$, the strongly nonlinear dependence indicates a large AH; AH vanishes at less negative $n$. Each curve in (A and B) is offset for clarity, and the dashed lines denote zero $\rho_{yx}$ signal. (**C**) The individual squares represent measured AH resistance, with color scale shown in the lower left corner. The circles represent data with no measureable AH.

Supplementary Materials for

# Tunable ferromagnetism at non-integer filling of a moiré superlattice

Guorui Chen[+], Aaron L. Sharpe[+], Eli J. Fox[+], Shaoxin Wang, Bosai Lyu, Lili Jiang, Hongyuan Li, Kenji Watanabe, Takashi Taniguchi, Michael F. Crommie, M. A. Kastner, Zhiwen Shi, David Goldhaber-Gordon[*], Yuanbo Zhang[*], Feng Wang[*]

[+]These authors contributed equally to this work.
*Corresponding author. Email: fengwang76@berkeley.edu, zhyb@fudan.edu.cn, goldhaber-gordon@stanford.edu

**This PDF file includes:**
Supplementary Text
Figs. S1 to S8
References and Notes

**Contents**

1. Gate map of the ABC-TLG/hBN device

2. Comparison of AH between $n = -n_0$ and $n = -2.3\ n_0$

3. $n$- and $D$-dependence of the strongest AH signals

4. Additional anomalous Hall data

5. Discussion on the origin of observed AH at non-integer filling

6. Data for the second device

**Supplementary Text**

1. Gate map of the ABC-TLG/hBN device

   The color plot of longitudinal resistance $R_{xx}$ as a function of $n/n_0$ and $D$ in Fig. 1F is calculated from the initially measured gate map of $R_{xx}$ as a function of $V_t$ and $V_b$ in Fig. S1. From the capacitance model and quantum Hall effect, the fully filled point at $n = \pm 4n_0$ corresponds to the carrier density of $n = \pm 2.1 \times 10^{12}$ cm$^{-2}$, from which the moiré period is estimated to be 15 nm and the twist angle between ABC-TLG and hBN is zero (*2*).

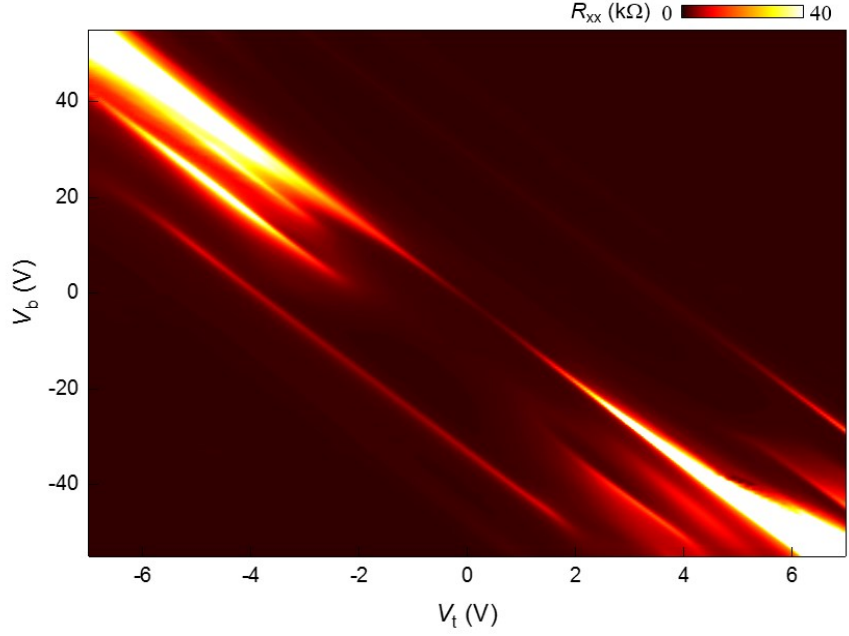

**Fig. S1. Gate map of ABC-TLG/hBN.** Color plot of $R_{xx}$ as a function of $V_t$ and $V_b$ measured at $T$ = 1.5 K. The data of Fig. 1F in the main text are created from these based on the relation between $n$, $D$ and $V_t$, $V_b$ discussed in the main text.

2. Comparison of AH between $n = -n_0$ and $n = -2.3\ n_0$

To compare the magnetic states in ABC-TLG/hBN, we plot $R_{yx}$ as a function of $B$ at $n = -n_0$ for the Chern insulating state and $n = -2.3n_0$ for the non-integer filling state at different temperatures side by side in Fig. S2. Here we select different $D$ so that the strongest AH is shown for both states. The AH of the non-integer filling magnetic state at $n = -2.3\ n_0$ has smaller amplitude than the AH of the Chern insulator state at $n = -n_0$, and they have opposite signs. The opposite sign AH signals might be an indication that the net valley polarizations are opposite for $n = -n_0$ and $n = -2.3\ n_0$. There are several jumps in the hysteretic loop at $n = -2.3\ n_0$ suggesting a complex magnetic domain structure. The AH signal at $n = -2.3\ n_0$ disappears at $T = 1.6$ K, lower than that of the Chern insulator state at $n = -n_0$, which persists to 3.5 K. In addition, the AH signal of the Chern insulator has a well quantized value at finite magnetic field and shows a quantum Hall dispersion governed by the Streda relation $n = CeB/h$ for $C = 2$. The AH at $n = -2.3\ n_0$, however, does not become quantized even at high magnetic field.

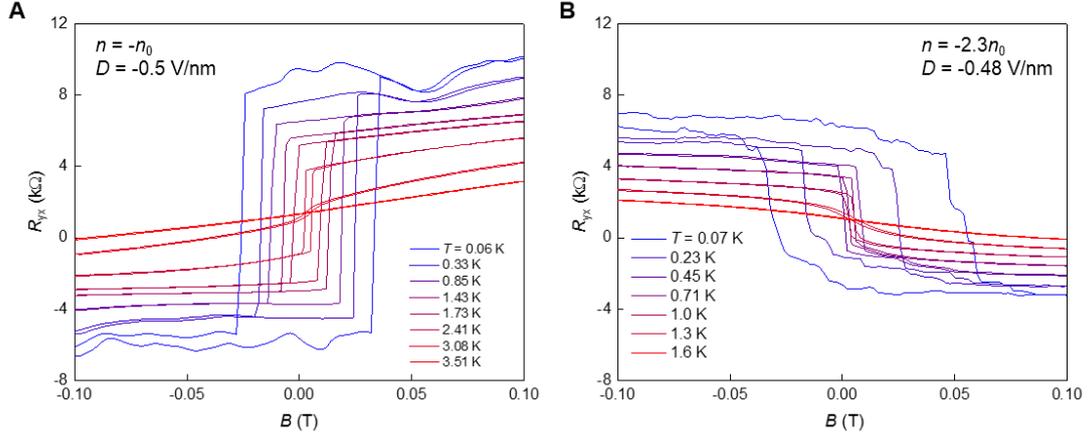

**Fig. S2. Ferromagnetism at $n = -n_0$ and $n = -2.3\, n_0$.** $R_{yx}$ as a function of $B$ field at different temperatures for (**A**) $n = -n_0$ and (**B**) $n = -2.3\, n_0$. The $D$ field is selected to be -0.5 V/nm in (A) and -0.48 V/nm in (B) to yield the largest $R_{yx}^{AH}$ for the same device.

3. *n*- and *D*-dependence of the strongest AH signals

Near $D = -0.48$ V/nm and $n = -2.3\, n_0$, where the AH signal is largest (see Fig. 4C in the main text), the longitudinal and Hall resistivities are quite sensitive to $D$ and $n$. Fig. S3A shows $\rho_{xx}$, $\rho_{yx}^{AH}$ and $B_c$, the coercive field, for densities near this point. Whereas there is a large peak in $\rho_{xx}$ at the integer filling of $n = -2\, n_0$, at most a very weak AH signal is detected there. On the other hand, the AH signal and the coercive field are maximum near $n = -2.3\, n_0$. Figure S3B shows that there is a similar sensitivity to $D$ in this region. Note that at this value of $D$ we do not see the sign change of AH discuss in the text, seen at less negative $D$.

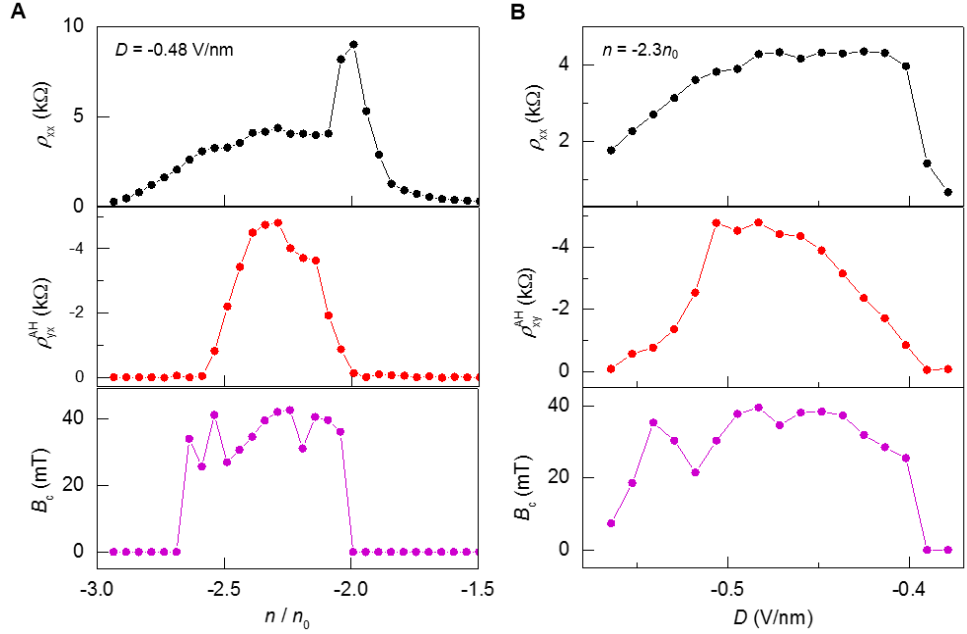

**Fig. S3.** *n*- **and** *D*-**dependence of ferromagnetism at non-integer filling.** (**A**) *n*- and (**B**) *D*-dependence of AH of the non-integer filling centered at $n = -2.3\, n_0$ and $D = -0.48$ V/nm at $T = 0.07$ K.

4. Additional anomalous Hall data

We present here a sampling of the many magnetic field sweeps taken to generate the data points in Fig. 4C in the main text. Figure S4 shows two values of *n*, which display sign changes as *D* is varied and two values of *D* that show sign changes as *n* is varied. The AH hysteretic loops at non-integer filling can be rather complicated at some points in the parameter space of *n* and *D*. Some loops, such as $D = -0.48$ V/nm in Fig. S4B and $n = -2.65\, n_0$ in Fig. S4D, show multiple intermediate jumps, and some hysteresis loops, such as that at $D = -0.48$ V/nm and $n = -2.5\, n_0$ in Fig. S4B and those at $n = -2.70\, n_0$ and $-2.61\, n_0$ at $D = -0.45$ V/nm in Fig. S4D, appear at non-zero magnetic fields, suggesting a complex domain structure.

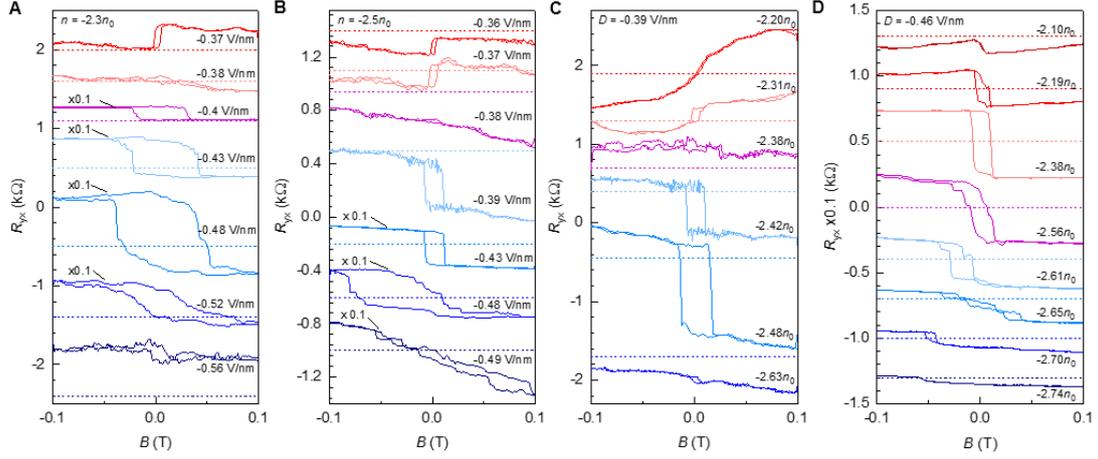

**Fig. S4. Additional $R_{yx}$ data.** $R_{yx}$ as a function of magnetic field at (**A**) $n = -2.3\ n_0$, and (**B**) $n = -2.5\ n_0$ for various $D$, and (**C**) $D = -0.39$ V/nm and (**D**) $D = -0.46$ V/nm for various $n$. Each curve is offset for clarity, and the dashed line with the same color as the data denotes $R_{yx} = 0$ for that data set. The dashed line is not located at the center of the corresponding AH loop due to the cross-talk from $R_{xx}$. The $R_{yx}$ signals for $D = -0.52$, $-0.48$, $-0.43$ and $-0.4$ V/nm in (A) and $D = -0.49$, $-0.48$ and $-0.43$ V/nm in (B) are manually multiplied by a factor of 0.1. Data of this kind are used to generate the data points in Fig. 4C of the main text.

5. Discussion on the origin of observed AH at non-integer filling

By plotting the behavior of the conductivity tensor $\sigma_{xx}$ and $\sigma_{yx}$, we can get more insight into the mechanism of the observed AH at non-integer filling. The AH effect can either be intrinsic, arising from Berry curvature of the filled bands, or extrinsic, resulting from scattering mechanisms. We suggest AH at the non-integer filling is intrinsic. First, the AH is only observed for the sign of the $D$ field which leads to the Chern band according to our band structure calculation, and not at the other sign which is predicted to lead to a topological trivial band. Second, skew scattering, one of the extrinsic mechanisms leading to AH in magnetic metals, gives rise to a linear relationship between $\sigma_{yx}$ and $\sigma_{xx}$. Figure S5A and B clearly show nonlinear relationships between $\sigma_{yx}$ and $\sigma_{xx}$ when either density or temperature are varied *(32)*. It is more challenging to rule out side jump scattering, another extrinsic mechanism. Both of these mechanisms, however, can be effective only in metallic materials and, as shown in Fig. 2B of the main text, the AH occurs where the resistance is quite high.

Furthermore the longitudinal conductivity increases with increasing temperature, as shown in Fig. S5C. Third, the largest anomalous Hall angle $\rho_{yx}/\rho_{xx}$ is close to unity, which is much larger than previously reported extrinsic or intrinsic AH materials ($\leq 0.2$) *(32,33)* except for Chern insulators *(5–7,34–36)*, which also suggests the intrinsic mechanism.

Fig. 5D shows $\sigma_{yx}$ and $\sigma_{xx}$ as functions of $n$ at $D = -0.48$ V/nm at $B = 0$ T. The large $\sigma_{yx}$ is accompanied with a drop of $\sigma_{xx}$, which is one would expect for a Chern insulator. We believe that the absence of quantization of the former and non-zero value of the latter probably results from the presence of domains.

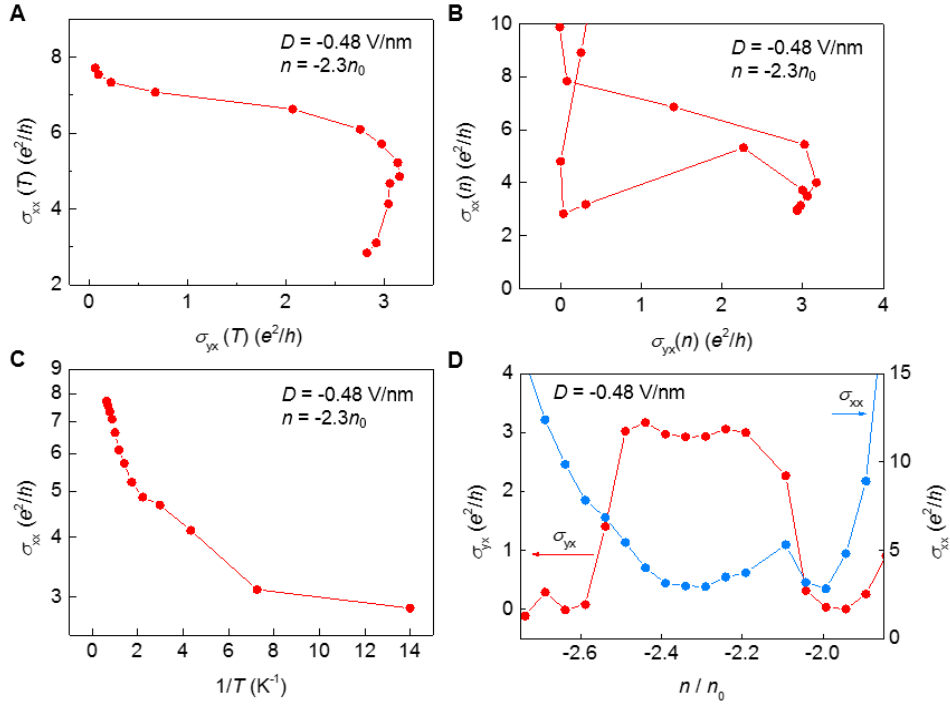

**Fig. S5. Behavior of the conductivity tensor at $B = 0$.** (**A**) The longitudinal conductivity $\sigma_{xx}$ is plotted parametrically against the Hall conductivity $\sigma_{yx}$ for a series of measurements at different temperatures with the density and $D$ field fixed at $n = -2.3\ n_0$ and $D = -0.48$ V/nm. (**B**) $\sigma_{xx}$ is plotted parametrically against $\sigma_{yx}$ for a series of measurements at different densities at $T = 0.07$ K, with $D = -0.48$ V/nm. (**C**) Arrhenius plot of $\sigma_{xx}$ on a log scale versus $1/T$. (**D**) $\sigma_{xx}$ and $\sigma_{yx}$ as functions of carrier density, from the same data as in (B), showing the emergence of a dip in $\sigma_{xx}$ accompanied by the $\sigma_{yx}$ maximum around $n = -2.3n_0$.

6. Data for the second device

The ferromagnetism at the non-integer filling is reproduced in the second ABC-TLG/hBN device. Fig. S6A and B shows the optical image and schematic side view of the second device. Figure S6C shows $R_{xx}$ as a function of $n$ and $D$ at $T = 5$ K. The ABC-TLG is aligned with the top hBN at zero twist angle. The Chern band is, therefore, at positive $D$ with weak resistance peaks at $n = -n_0$ and $-2 n_0$, and the topological trivial band is at negative $D$ with Mott strong insulating resistance peaks at $n = -n_0$ and $-2 n_0$.

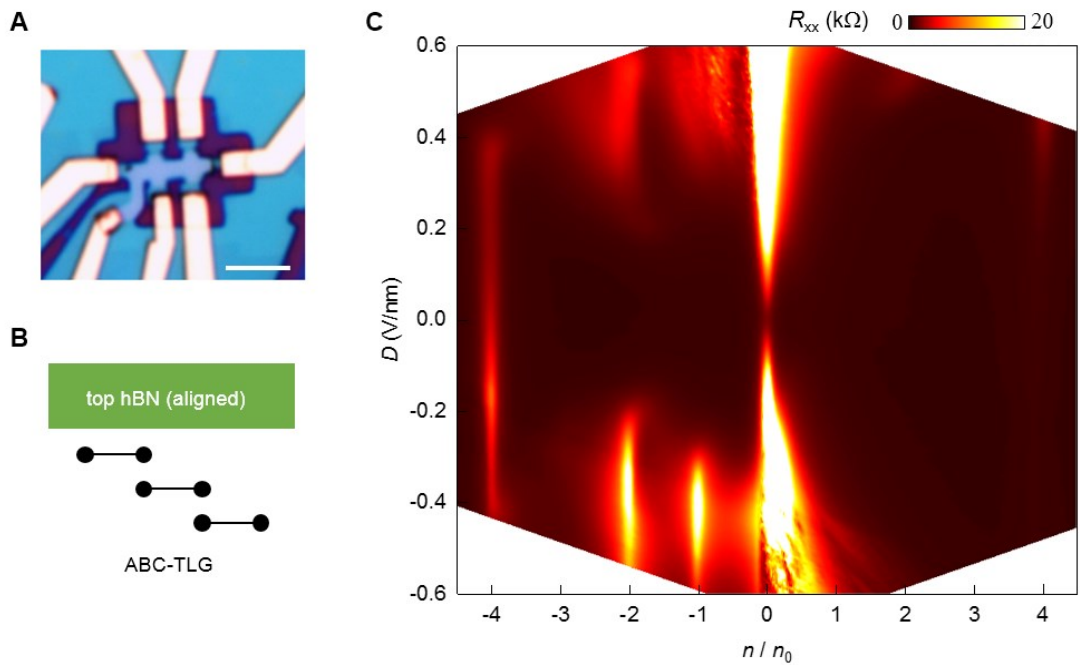

**Fig. S6. Basic characterization of the second device.** (**A**) Optical image the second device. Scale bar: 5 μm. (**B**) The illustration of the unit cell of ABC-TLG aligned with the top hBN layer. (**C**) The $R_{xx}$ as a function of carrier density and vertical displacement field at $T = 5.2$ K.

Based on the knowledge from the first device discussed in the main text, we focus on the valence minibands on the non-trivial side of $D$. In Fig. S7A we indicate values of $n$ and $D$ for which $R_{yx}$ had been measured using red-, blue- and gray-filled squares, corresponding to positive, negative and zero AH, respectively. Fig. S7B shows some typical AH hysteresis loops at $D = 0.37$ V/nm for various $n$. These show the sign change of AH discussed in the main text with reference to Fig. 4. Figures S7C, D and E show the evolution of the AH with carrier density. In Fig. S7C we plot $R_{xy}$ for as a function of *n* for the two directions of field sweep, which shows the hysteresis. Taking the difference of these two sweeps shows the size and sign of the AH as displayed in Fig. S7D. Figure S7E shows how the coercive field varies with *n*.

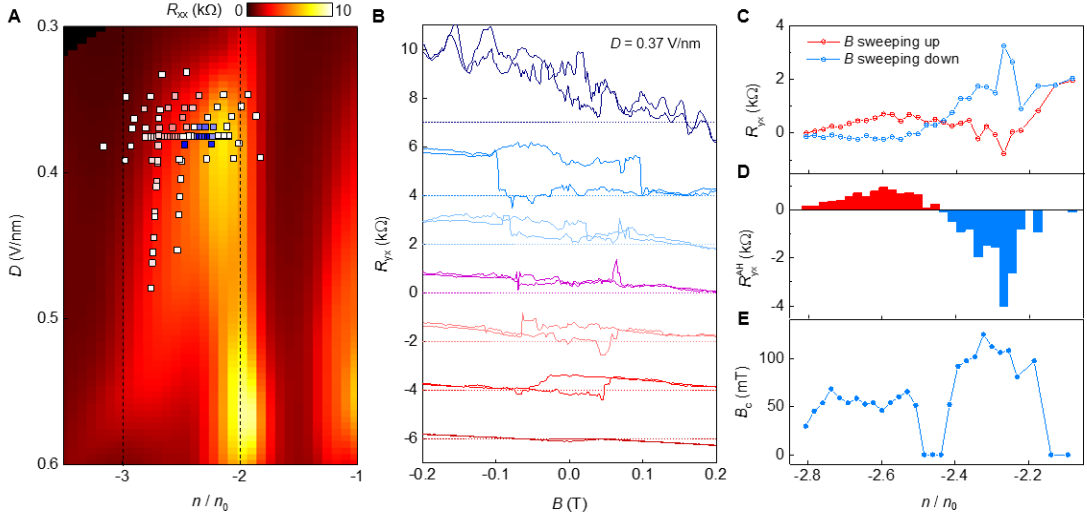

**Fig. S7. Non-integer filling ferromagnetism in the second device.** (**A**) The background color scale gives $R_{xx}$ as a function of $n/n_0$ and $D$ for the non-trivial side. The red- and blue-filled squares correspond to positive and negative AH where blue-white-red correspond to $R_{yx}^{AH}$ = -2 kΩ to 0 to 2kΩ. (**B**) The evolution of AH hysteresis at different doping for $D$ = 0.37 V/nm. From top to bottom: $n$ = -2.14$n_0$, -2.37$n_0$, -2.48$n_0$, -2.53$n_0$, -2.62$n_0$, -2.81$n_0$, manual offset of 6 kΩ, 4 kΩ, 2 kΩ, 0, -2 kΩ, -4 kΩ, -6 kΩ. (**C**) $R_{yx}$ at $B$ = 0 T for up and down sweeps of $B$ field as a function of carrier density for $D$ = 0.36 V/nm. (**D** and **E**) The $\rho_{yx}^{AH}$ and $B_c$ as a function of carrier density for $D$ = 0.36 V/nm.

Fig. S8 compares the *n*- and *D*-dependent longitudinal resistivity and AH behavior in device 1 (discussed in the main text) and device 2. The qualitative behaviors are similar: (1) The ferromagnetic state emerges in the doping range of -2.2 to -2.8 $n_0$. (2) Both negative and positive ferromagnetic hysteresis are present and they depend sensitively on *D* and *n*. However, there are significant differences in the detailed ferromagnetic phase diagram. The magnetic states are more limited in *D* range, 0.35 to 0.4 V/nm, in device 2, compared to a range of about -0.35 to -0.55 V/nm for device 1. The $R_{yx}^{AH}$ in device 2 is also smaller than the first device, with the largest AH at ~ 2 kΩ. In addition, the sign of the AH in device 2 is mainly determined by the doping, where positive AH emerges at larger doping and negative one at smaller doping. On the other hand, the sign of the AH in device 1 depends more on the displacement field $D$. The origin of these differences between the two devices is not known. It could arise from small differences in the layer stacking details, such as moiré superlattice inhomogeneity, between ABC-TLG and hBN layers.

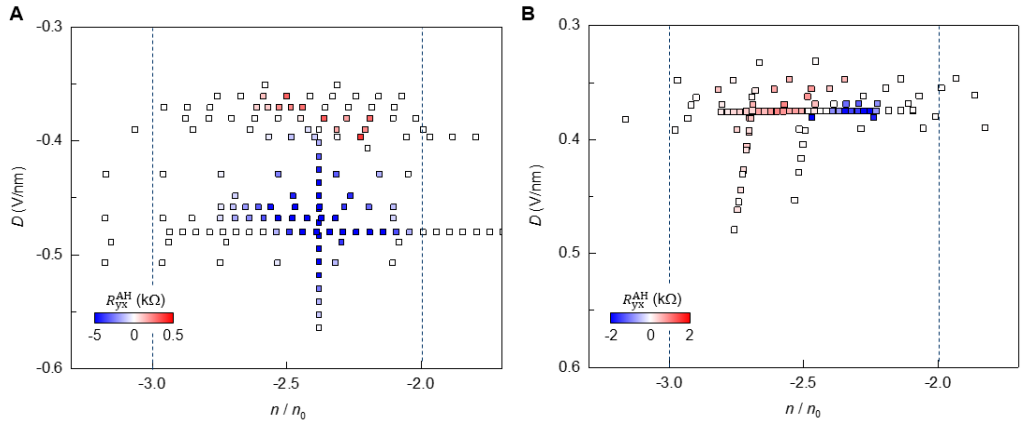

**Fig. S8**. **Ferromagnetism summary of non-integer filling for two devices.** Observed magnetic states of device 1 (**A**) and device 2 (**B**). The measured $R_{yx}^{AH}$ values are represented as individual sqaures on the color plot of $R_{xx}$ as a function of $n$ and $D$. The color of the points are linked to the $R_{yx}^{AH}$ values using a blue-white-red scale corresponding to: -5 kΩ to 0 to 0.5kΩ for (A), -2 kΩ to 0 to 2 kΩ for (B).